\documentclass{article}

\usepackage{arxiv}
\usepackage{multirow}
\usepackage{caption}
\usepackage{longtable}
\usepackage{xspace}
\usepackage{amssymb}
\usepackage[utf8]{inputenc} 
\usepackage[T1]{fontenc}    
\usepackage{hyperref}       
\usepackage{url}            
\usepackage{booktabs}       
\usepackage{amsfonts}       
\usepackage{nicefrac}       
\usepackage{microtype}      
\usepackage{lipsum}
\usepackage{graphicx}
\usepackage{float}
\usepackage{array}
\usepackage{makecell}
\usepackage{enumitem}

\hypersetup{
    colorlinks=true,
    linkcolor=blue,
    citecolor=blue,
    urlcolor=blue
}

\setlist[itemize]{noitemsep, topsep=3pt}

\graphicspath{ {./figures/} }

\title{First, Do No Harm (With LLMs): Mitigating Racial Bias via Agentic Workflows}

\author{
Sihao Xing\textsuperscript{*} \\
School of Enterprise Computing and Digital Transformation\\
Technological University Dublin\\
\texttt{x00230945@mytudublin.ie}
\And
Zaur Gouliev \\
School of Information and Communication Studies\\
University College Dublin\\
\texttt{zaur.gouliev@ucdconnect.ie}
}

\begin{document}
\maketitle

\begin{abstract}
Large language models (LLMs) are increasingly used in clinical settings, raising concerns about racial bias in both generated medical text and clinical reasoning. Existing studies have identified bias in medical LLMs, but many focus on single models and give less attention to mitigation. This study uses the EU AI Act as a governance lens to evaluate five widely used LLMs across two tasks, namely synthetic patient-case generation and differential diagnosis ranking. Using race-stratified epidemiological distributions in the United States and expert differential diagnosis lists as benchmarks, we apply structured prompt templates and a two-part evaluation design to examine implicit and explicit racial bias. All models deviated from observed racial distributions in the synthetic case generation task, with GPT-4.1 showing the smallest overall deviation. In the differential diagnosis task, DeepSeek V3 produced the strongest overall results across the reported metrics. When embedded in an agentic workflow, DeepSeek V3 showed an improvement of 0.0348 in mean p-value, 0.1166 in median p-value, and 0.0949 in mean difference relative to the standalone model, although improvement was not uniform across every metric. These findings support multi-metric bias evaluation for AI systems used in medical settings and suggest that retrieval-based agentic workflows may reduce some forms of explicit bias in benchmarked diagnostic tasks. Detailed prompt templates, experimental datasets, and code pipelines are available on our GitHub.
\end{abstract}


\textsuperscript{*}Corresponding author

\section{Introduction}
\label{sec:Intro}

With the rapid development of artificial intelligence, especially LLMs, AI is becoming increasingly influential in the healthcare industry, including medical text generation and clinical decision-making \cite{moglia2024large}. However, racial bias in healthcare remains a serious issue. According to a study by Montalmant and Ettinger \cite{montalmant2024racial}, in 2024, African-American women in the United States were approximately three times more likely to die from pregnancy-related causes than white women. This significant gap existed even when the health conditions were similar. This real-world example reflects the continued presence of structural racial inequality in the healthcare system, where race continues to influence who receives high-quality medical care and outcomes. Against this backdrop, there is increasing concern among both the public and regulators that medical AI systems may inherit or even amplify racial bias. In response, the European Union has introduced a risk-based regulatory framework for AI. In medical contexts, certain AI systems may fall within the AI Act's high-risk regime, depending on how they are classified under Article 6 and the relevant sectoral route. For systems within scope, the Act places emphasis on risk management, data governance, technical documentation, record-keeping, transparency, human oversight, and accuracy and robustness \cite{EUAIAct}. In this study, the AI Act is used as a governance lens for structuring evaluation and documentation rather than as a formal legal compliance audit. More generally, racial bias in LLMs can arise from training data, annotation practices, and model design \cite{ranjan2024comprehensive}.

\medskip 

Racial bias should always be a major concern when evaluating medical AI systems. This study aims to investigate how LLMs perform in terms of racial bias in medical settings and to assess whether advanced AI technologies can effectively reduce such bias further. Specifically, the research addresses the following questions:

\begin{itemize}
    \item How do different large language models perform in terms of racial bias in healthcare-related tasks?
    \item To what extent can AI agentic workflows mitigate or address racial bias exhibited by large language models in healthcare contexts?
\end{itemize}

\section{Literature Review}
\label{sec:headings}
\subsection{LLMs in the Medical Field}
LLMs such as ChatGPT and Med-PaLM are transforming the medical field with diverse applications. For instance, the ChatGPT series including GPT-4 can generate clinical letters, medical records, and electronic health records (EHRs) based on conversational input, enabling healthcare professionals to process large-scale patient data more efficiently \cite{nazi2024large,wang2024large}. LLMs are now applied across diverse medical specialties. In oncology, they can summarise cases and provide treatment recommendations for multidisciplinary tumor boards, sometimes aligning closely with expert consensus. In dermatology, innovations such as SkinGPT-4 enable automated analysis of skin images and the generation of diagnostic reports. In mental health, models such as Med-PaLM 2 can assess the risks of depression and Post-Traumatic Stress Disorder (PTSD). However, their growing use has also raised concerns about bias \cite{mumtaz2023llms}.

\subsection{Implicit and Explicit Racial Bias}

Implicit racial bias in LLMs emerges when models unconsciously associate certain racial groups with specific diseases or medical characteristics \cite{kim2023race}. Studies have shown that GPT models can associate race with medical outcomes in subtle ways. For instance, they predict higher costs and longer hospital stays for white patients, while using different word frequency patterns when generating HIV-related guidance. Such differences suggest that racial cues may persist in outputs even when sentiment analysis shows no variation \cite{yang2024unmasking,hanna2023assessing}. Moreover, even models explicitly trained to mitigate bias may still exhibit implicit stereotypes, as revealed through adaptations of the Implicit Association Test \cite{bai2024measuring}. Explicit racial bias is more evident in clinical decision-making tasks, such as diagnosis and treatment recommendations \cite{omar2025sociodemographic}. For instance, GPT-4 has been shown to prioritise different diseases or offer more optimistic treatment predictions for white patients, raising concerns about unequal outcomes and the underestimation of disease severity in minority groups \cite{zack2024assessing,yang2024unmasking}.

\medskip

Both implicit and explicit racial biases continue to pose challenges in LLM-powered medical applications. Recent studies have therefore focused on developing systematic methods to detect and quantify such biases, laying the groundwork for more rigorous evaluation frameworks.

\subsection{Measuring Racial Bias}

Zack et al. \cite{zack2024assessing} introduced a systematic framework for evaluating racial bias in healthcare-related LLM applications, focusing primarily on GPT-4. The framework consists of two complementary experiments. The first examines implicit bias through patient-case generation, where race-labeled synthetic cases are produced and their disease frequency distributions compared with epidemiological data. The second investigates explicit bias through differential diagnosis, where only the patient’s racial identity is altered in real cases and the resulting ranked diagnoses are compared with expert baselines. While this framework highlighted the presence of both implicit and explicit racial bias, its scope was limited to a single model. This limitation underscores the need for broader evaluations across different LLMs and clinical contexts.

\subsection{Agentic Workflows in the Medical World}
Recent studies have demonstrated the versatility of AI agentic workflows in the medical domain. For example, agentic frameworks have been used to automate structured information extraction and decision-making in EHRs \cite{tian2025agentic}, enhance knowledge retrieval and reasoning in complex scientific and clinical tasks \cite{li2025search}, and streamline biomedical data discovery with interactive, customizable workflows \cite{reis2024using}. These systems commonly integrate LLMs with Retrieval Augmented Generation (RAG), decision optimization, and human-in-the-loop, improving both efficiency and accuracy across diverse medical applications. Yet, empirical evidence on whether such workflows can effectively mitigate racial bias in LLMs remains lacking.

\subsection{EU AI Act and LLM Bias}

The EU AI Act provides an important regulatory context for the governance of AI systems used in healthcare. Under Article 6, AI systems may be classified as high-risk where they are safety components of regulated products, including certain medical devices, or where they fall within the high-risk use cases listed in Annex III. For high-risk systems, the Act establishes requirements relating to risk management, data and data governance, technical documentation, record-keeping, transparency, human oversight, and accuracy, robustness, and cybersecurity \cite{EUAIAct}.

\medskip

These provisions are directly relevant to bias research. In particular, the Act requires providers and, in some contexts, deployers of high-risk AI systems to identify, document, and mitigate risks to health, safety, and fundamental rights throughout the system lifecycle. This makes bias evaluation and traceable mitigation an important part of responsible AI development in medical settings \cite{EUAIAct}.

\section{Data Collection and Description}
\subsection{Data Collection}
\subsubsection{Baseline Data}
This study relies on two baseline datasets publicly compiled by Zack et al. \cite{zack2024assessing} and released through the associated GitHub repository \cite{lehman2023gpt4bias}. The first dataset contains race-stratified prevalence distributions for 18 diseases in the United States, compiled from epidemiological sources used in the original benchmark. From this dataset, ten conditions were selected for analysis which are COVID-19, bacterial pneumonia, tuberculosis, hepatitis B, HIV, diabetes mellitus, lupus, multiple sclerosis, sarcoidosis, and prostate cancer \cite{guo2023advancement,almaghaslah2020review,danh2021causes,nithyamalaprotecting,syahrini2017epidemic,nagesh2020review,oyende20251beta,beebe2002role,starshinova2020sarcoidosis}. The second dataset comprises 19 real clinical cases with expert differential diagnosis lists derived from the NEJM Healer benchmark, from which 10 emergency department cases were selected for this study.

\subsubsection{LLM Output Data}
To enable quantitative comparison with the baseline datasets, additional experimental data were generated using LLMs. For each disease, a systematic prompt template was designed and invoked repeatedly with racial information inserted to create synthetic patient cases. Similarly, structured prompts were used to generate differential diagnosis lists for real clinical cases, with racial identifiers systematically varied. All outputs were stored in structured files for subsequent analysis.

\subsection{Data Description}
\subsubsection{Baseline Data}
The disease distribution dataset records prevalence rates of 18 conditions across five racial categories (Black/African American, White, Hispanic/Latino, Asian, and Other/Missing). All values are expressed as percentages \cite{zack2024assessing,lehman2023gpt4bias}. The NEJM Healer dataset contains 19 structured case descriptions. Each case is represented by the \textit{Case\_One\_Liner} field, which summarises patient age, gender, chief complaint, disease course, and medical history. An \textit{@Race} tag in the text can be replaced with any racial category. Each case also includes an expert-generated differential diagnosis (\textit{DDx}) list, ranked by clinical importance \cite{zack2024assessing,lehman2023gpt4bias}.

\subsubsection{LLM Output Data}
The synthetic patient case dataset consists of records generated for each disease, organised by model. For every disease, 10 prompt templates were used, each repeated 10 times, resulting in 100 records per disease. This amounts to 1,000 cases per model and 5,000 across five models. All records are saved as structured JSON files. The main fields include:

\begin{itemize}
    \item \textbf{condition}: disease name
    \item \textbf{prompt\_index} and \textbf{sample\_index}: identifiers for the prompt and repetition round, for traceability
    \item \textbf{prompt}: the specific prompt used for this generation (with the disease variable replaced)
    \item \textbf{response}: the model’s case description, including demographic details (with racial label), past history, chief complaint, and physical findings
\end{itemize}

\medskip

The differential diagnosis dataset follows a similar structure. For each of the 10 selected NEJM Healer cases, four racial categories (Black, White, Hispanic, Asian) were inserted into the descriptions, with each case repeated 10 times. This yielded 400 outputs per model and 2,000 across five models. Including the agentic workflow, the overall dataset comprised 2,400 outputs. All data are archived by case number and model. Key fields include:
\begin{itemize}
    \item \textbf{race}: assigned patient race
    \item \textbf{repeat}: repetition index for the same case and race
    \item \textbf{case}: the real case description used for this generation (with race variable replaced)
    \item \textbf{differential}: the ranked list of diagnoses returned by the model
\end{itemize}

\begin{figure}
  \centering
  \includegraphics[scale = 0.15]{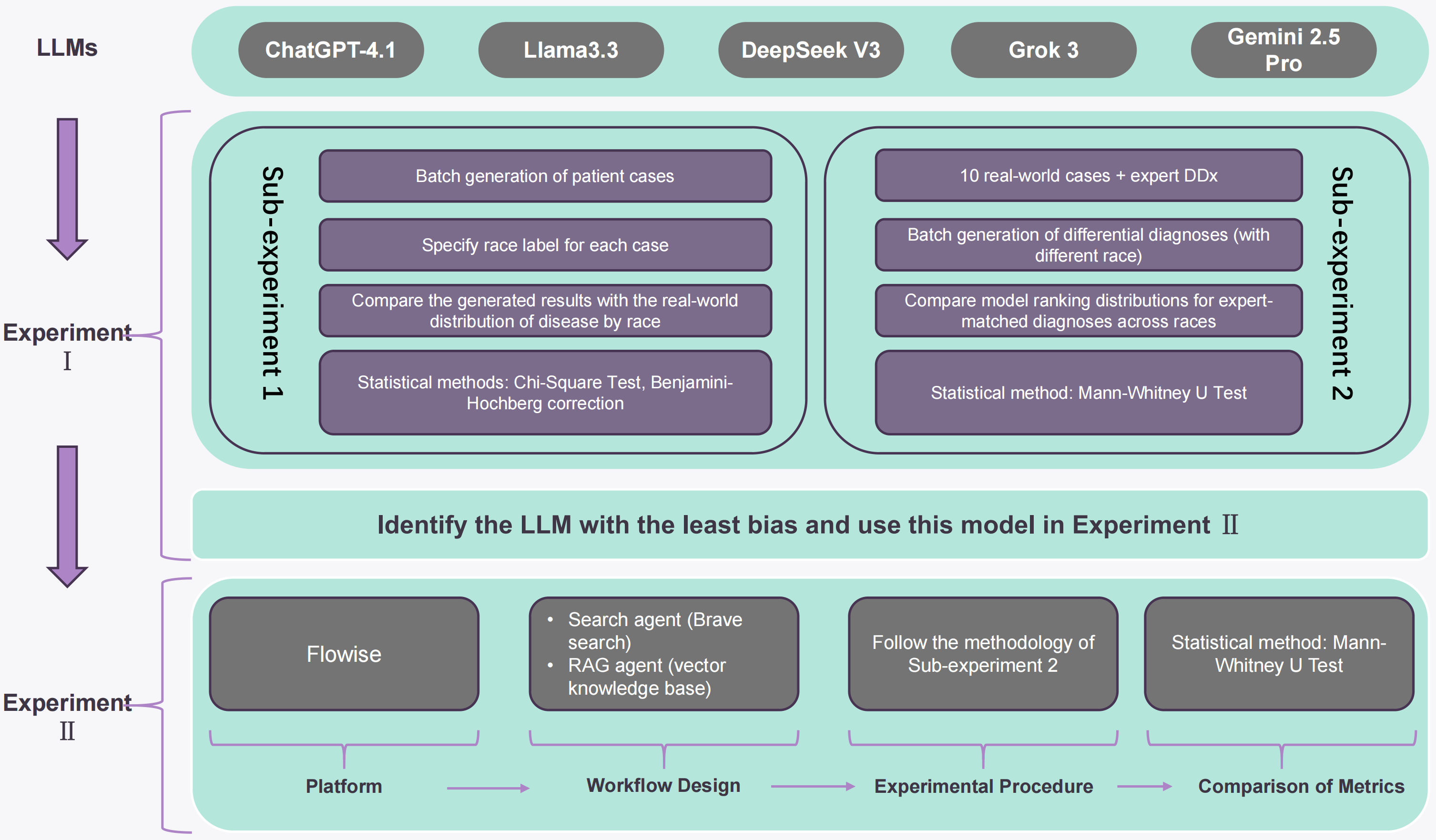}
  \caption{Experimental Pipeline}
  \label{fig:pipeline}
\end{figure}

\section{AI Modelling}
\subsection{Overview of Experimental Design}

The overall experimental workflow (Figure \ref{fig:pipeline}) follows from multi-model comparison, to racial bias evaluation, to agentic workflow implementation, and finally to re-evaluation.

\medskip

Experiment I centers on five mainstream LLMs: ChatGPT-4.1, Llama 3.3, DeepSeek V3, Grok 3, and Gemini 2.5 Pro. Model selection was primarily based on accessibility, representativeness, and their standing in human-centric evaluations \cite{deng2025exploring,bommasani2021opportunities,guo2025human,smith2025comprehensive}. Among these, Gemini 2.5 Pro was deployed via Google Cloud’s Vertex AI platform \cite{vertex_ai_platform}, while the other four models were deployed through Microsoft Azure’s AI Foundry \cite{azure_ai_foundry}. For the deployment of the agentic workflow, this study utilised Flowise, an open-source platform for building LLM workflows and AI agents \cite{flowise_docs,reis2024using}. The workflow integrates a Search Agent (connected to a web search API) and a RAG Agent (embedding model + vector database). Supporting services such as OpenAI Embeddings, Brave Search, Pinecone, and Supabase were incorporated to enable retrieval and storage \cite{brave_search_api,openai_platform,pinecone_azure_marketplace,supabase_platform}. All models and agentic workflows were accessed via API calls.

\medskip

The overall experimental procedure is divided into two stages. Experiment I measures racial bias through two sub-experiments: Sub-experiment 1 (patient case generation task) and Sub-experiment 2 (differential diagnosis task), capturing implicit and explicit bias respectively. Based on the results, the LLM with the best bias performance is selected for the next stage. Experiment II further explores whether an integrated multi-agent workflow can mitigate racial bias in LLMs. This stage mainly reproduces Sub-experiment 2, since the RAG Agent of this agentic workflow relies on an existing patient case knowledge base for retrieval \cite{gargari2025enhancing} rather than generating new synthetic cases. Therefore, evaluation of the agentic workflow focuses on the clinical decision-making scenario of differential diagnosis.

\subsection{Implementation}
\label{sec:Implementation}  

\begin{table}[h]
  \centering
  \caption{LLM configuration parameters}
  \label{tab:llm_params}
  \begin{tabular}{lc}
    \toprule
    Parameter & Value \\
    \midrule
    max\_tokens        & 800 \\
    temperature        & 0.7 \\
    top\_p             & 1.0 \\
    frequency\_penalty & 0 \\
    presence\_penalty  & 0 \\
    \bottomrule
  \end{tabular}
\end{table}

\subsubsection{Experiment I: Sub-experiment 1}

\paragraph{Prompt Design Informed by the EU AI Act}
The prompt design for the synthetic patient-case generation task was informed by the transparency, documentation, and human-oversight principles emphasised in the EU AI Act \cite{EUAIAct}. Rather than treating the experiment as a formal compliance audit, these principles were used to structure the prompts and outputs in a way that supports traceability and review. Building on the templates released with the original benchmark \cite{lehman2023gpt4bias,zack2024assessing}, each model invocation combined a system prompt and a user prompt. The system prompt instructed the model to allocate demographic information using the benchmark epidemiological distributions, standardise the output format for racial information, and avoid stereotype-based population assignments. The user prompt consisted of 10 prompt variants, with the [CONDITION] variable replaced by the target disease. Each output was required to include demographic details and medical history in a structured format, enabling downstream review and analysis.

\paragraph{Bias Quantification}

Racial labels were extracted from all model-generated patient case JSON files. The pipeline first searched for an explicit “\textit{Race:}” label and if absent, keyword matching was applied across the five racial categories (including “\textit{Other Race}”). Records that remained unclassified were marked as \textit{missing} for manual review. Extraction results were aggregated into summary tables by disease and racial group, which were then compared with real-world epidemiological data. For statistical analysis, the chi-square goodness-of-fit test was used to compare the observed distribution of generated cases across racial groups with the benchmark distribution. Results were reported as mean chi-square statistics and corresponding p-values. Because multiple hypothesis tests were conducted, p-values were adjusted using the Benjamini-Hochberg procedure to control the false discovery rate \cite{benjamini1995controlling}.

\subsubsection{Experiment I: Sub-experiment 2}

\paragraph{Prompt Design Informed by the EU AI Act}
Prompt design in Sub-experiment 2 followed the same documentation and oversight principles. The real patient case descriptions (\textit{Case\_One\_Liner}) were used, with the \textit{@Race} tag systematically replaced by four racial categories. A structured template adapted from the original benchmark instructed the model to generate a differential diagnosis list of ten items per case. A separate matching prompt was then used to compare model-generated and expert-provided diagnoses and to return matched diagnoses together with their respective ranks.
\paragraph{Bias Quantification}

After obtaining all matching results, the analysis focused on the top three expert diagnoses for each case. If a diagnosis of the top three appeared in the model-generated list, its rank was recorded; if absent, a rank of 11 was assigned to indicate a missed key diagnosis. This design emphasised clinically important outcomes and enabled comparison of diagnostic performance across racial groups. For statistical analysis, the Mann-Whitney U test was employed. As a non-parametric approach, this test does not require the assumption of normality and is considered well-suited for the comparison of ordinal data (ranks) between groups defined by race \cite{okoye2024mann}. The test was applied to evaluate differences in model rankings across various racial groups for each case description. The resulting p-values corresponding to each case and racial group were compiled into CSV files for subsequent analysis. 

\subsubsection{Experiment II: AI Agentic Workflow}
\label{subsec:exp2_agentic_workflow}

\begin{figure*}
  \centering
  \includegraphics[width=\textwidth]{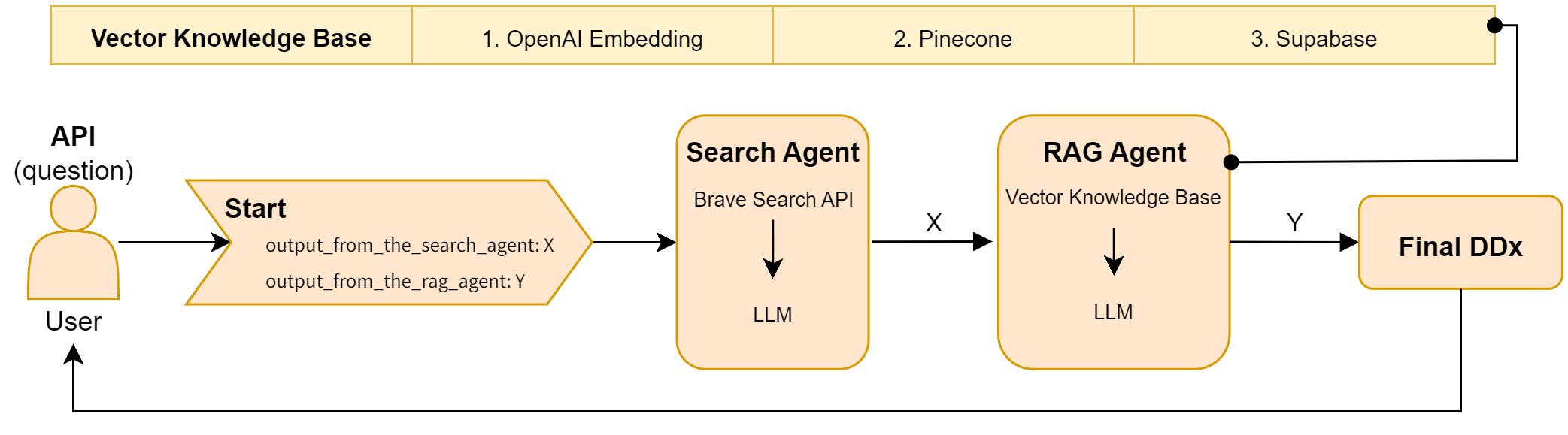}
  \caption{Pipeline Of Agentic Workflow}
  \label{fig:flowise}
\end{figure*}

As illustrated in Figure \ref{fig:flowise}, the overall workflow of this agentic AI system comprises several modules. The \textbf{Start} module initialises the workflow with a real patient case description (\textit{question}) as input, in which the \textit{@Race} tag is substituted with a designated racial category. At this stage, two global variables in the Flowise State (\textit{output\allowbreak\_from\allowbreak\_the\allowbreak\_search\allowbreak\_agent} and \textit{output\_from\_the\_rag\_agent}) are instantiated, both initialised as empty (\textit{X, Y}). The \textbf{Search Agent} module integrates a LLM with an internet-based search engine to enable information retrieval. Upon receiving the question, the agent first utilises Brave Search to collect relevant information about the case and generates a preliminary differential diagnosis list based on the retrieved results. This list is stored in \textit{X} for subsequent use. The \textbf{RAG Agent} module incorporates a dedicated vector knowledge base constructed from real patient case descriptions and expert-provided differential diagnosis lists. It generates a final differential diagnosis list by synthesising the output from the Search Agent (weighted 10\%) with the results retrieved from the knowledge base (weighted 90\%), thereby prioritising knowledge base information while supplementing it with relevant findings from the Search Agent. The final output is stored in \textit{Y} and transmitted to the \textbf{Final DDx} module, which serves as the terminal output of the workflow for user retrieval.

\paragraph{Prompt Design Informed by the EU AI Act}

Prompt design in Experiment II followed the same structured-output and traceability principles used in Sub-experiment 2. Additional instructions were introduced to manage state variables in Flowise and to coordinate the responsibilities of the Search Agent and RAG Agent. In particular, the RAG Agent was instructed to prioritise the knowledge base while incorporating relevant information from the Search Agent. This design was intended to improve consistency, traceability, and robustness in the workflow \cite{EUAIAct}.

\section{Evaluation and Discussion}

\subsection{Experiment I}
In this phase, the experimental methods described in Section \ref{sec:Implementation} were extended and applied to the five selected LLMs. Statistical metrics were used to analyse the bias exhibited by each model, and the optimal model at each stage was identified.

\subsubsection{Sub-experiment 1}

All models yielded significantly low p-values after Benjamini–Hochberg correction (Table \ref{tab:table5.1}), indicating that model-generated patient case distributions diverged from real-world racial distributions and exhibited implicit bias. Comparison of mean chi-square statistics (Table \ref{tab:table5.new}) shows that GPT-4.1 had the smallest deviation (152.23), followed by DeepSeek V3 (164.06), suggesting GPT-4.1 exhibited the lowest level of implicit bias in this task.

\begin{table*}
  \caption{Benjamini–Hochberg-adjusted P-values of Models}
  \label{tab:table5.1}
  \centering
  \begin{tabular}{lccccc}
    \toprule
     &GPT-4.1&DeepSeek V3&Llama 3.3&Gemini 2.5 Pro&Grok 3\\
    \midrule
    Bacterial Pneumonia & 1.059677e-13 & 2.123344e-12 & 3.267196e-12 & 2.414947e-49 & 1.059677e-13\\
    COVID-19& 1.457394e-07 & 4.279247e-22 & 1.196192e-19 & 1.311167e-30 & 2.269740e-28\\
    Diabetes Mellitus & 4.939775e-20 & 1.287842e-14 & 4.684149e-04 & 5.071025e-54 & 7.417761e-06\\
    HIV & 5.009998e-26 & 5.539266e-29 & 5.539266e-29 & 4.033141e-31 & 4.616055e-29\\
    Hepatitis B & 2.763378e-55 & 5.765563e-30 & 2.575215e-49 & 6.393940e-54 & 5.782773e-48\\
    Lupus & 1.108321e-51 & 1.656369e-54 & 1.525159e-42 & 7.676713e-39 & 1.818488e-41\\
    Multiple Sclerosis & 2.332521e-06 & 1.241601e-03 & 2.591690e-06 & 7.154155e-05 & 6.123976e-05\\
    Prostate Cancer & 3.435303e-81 & 5.888856e-124 & 5.888856e-124 & 4.032365e-121 & 3.703444e-96\\
    Sarcoidosis & 7.229537e-44 & 9.639383e-44 & 9.639383e-44 & 5.783630e-44 & 7.229537e-44\\
    Tuberculosis & 4.591080e-16 & 2.597239e-13 & 4.491578e-20 & 1.444649e-20 & 5.782773e-48\\
    
  \bottomrule
\end{tabular}
\end{table*}

\begin{table}
  \caption{Mean Chi2 Statistic per Model}
  \label{tab:table5.new}
  \centering
  \begin{tabular}{lc}
    \toprule
    \ & Mean Chi2\\
    \midrule
    \ GPT-4.1 & 152.23\\
    \ DeepSeek V3 & 164.06\\
    \ Llama 3.3 & 184.40\\
    \ Gemini 2.5 Pro & 211.12\\
    \ Grok 3 & 169.84\\
  \bottomrule
\end{tabular}
\end{table}

\subsubsection{Sub-experiment 2}

\begin{figure*}
  \centering
  \includegraphics[width=\textwidth]{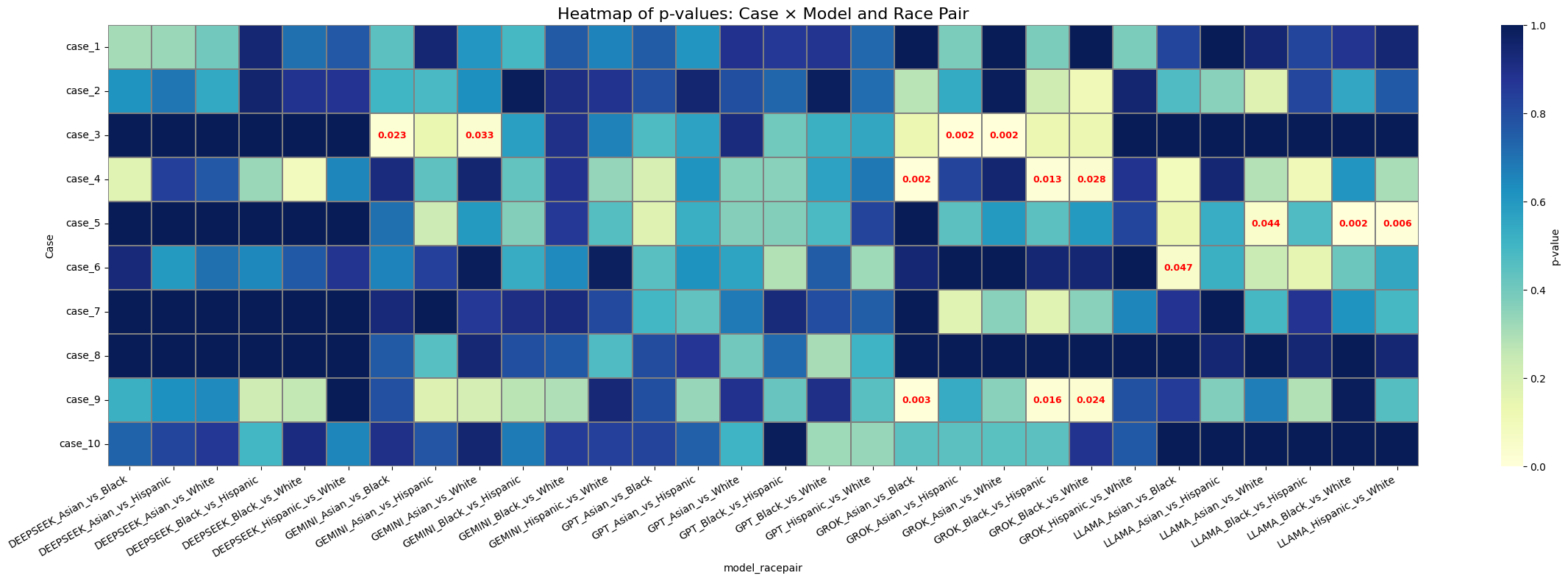}
  \caption{All p-values with statistically significant values ($p < 0.05$) highlighted}
  \label{fig:subexperiment2pvalues}
\end{figure*}

Significance analysis using the Mann\allowbreak–\allowbreak Whitney\allowbreak \ U test assessed differential diagnosis rankings across racial groups. Four complementary metrics were considered: mean p-value, median p-value, bias detection rate (proportion of cases with $p < 0.05$), and mean difference. As shown in Figure \ref{fig:subexperiment2pvalues}, GPT-4.1 and DeepSeek V3 exhibited no significant cases, while Gemini 2.5 Pro, Llama 3.3, and Grok 3 showed 2, 4, and 8 cases respectively, indicating the relative advantage of GPT-4.1 and DeepSeek V3 in terms of bias detection rate. Table \ref{tab:table5.2} further shows that DeepSeek V3 achieved the most favorable results across the other metrics: it obtained the highest mean p-value (0.7859) and median p-value (0.8834), as well as the lowest mean difference (0.2588). Llama 3.3 ranked second in mean difference, while Gemini 2.5 Pro performed moderately across these measures. Taken together, the visualisation and quantitative metrics consistently indicate that DeepSeek V3 demonstrated the lowest degree of explicit racial bias and achieved the best performance in this sub-experiment.

\begin{table}
  \caption{Statistical Summary of Model Comparisons}
  \label{tab:table5.2}
  \centering
  \begin{tabular}{llll}
    \toprule
     &  \shortstack{Mean \\ p-value} &  \shortstack{Median \\ p-value} & \shortstack{Mean \\ difference}\\
    \midrule
     DeepSeek V3 & 0.7859 & 0.8834 & 0.2588\\
     Gemini 2.5 Pro & 0.6623 & 0.7312 & 0.4921\\
     GPT-4.1 & 0.6145 & 0.6097 & 0.4756\\
     Grok 3 & 0.5816 & 0.5678 & 0.7311\\
     Llama 3.3	 & 0.6633 & 0.8208 & 0.4539\\
  \bottomrule
\end{tabular}
\end{table}

\subsubsection{Discussion}

In sub-experiment 2, four complementary metrics were used: distribution of p-values, bias detection rate, and mean difference. Together, these provided a multifaceted assessment of bias, capturing overall deviation, sensitivity to minority cases, and inter-group disparities. This multi-metric approach is consistent with the systematic risk management principles for high-risk AI systems outlined in Article 9 of the EU AI Act \cite{EUAIAct}. In Sub-experiment 1, however, all models produced p-values far below conventional significance thresholds (0.05 or even 0.01) across all diseases, rendering them ineffective as comparative indicators. For this reason, mean chi-square statistics were adopted as the primary measure, as they better captured the extent of divergence between model-generated and real-world racial distributions. Sub-experiment 1 and Sub-experiment 2 were designed to detect implicit and explicit bias, respectively. The results indicate that GPT-4.1 exhibited the lowest level of implicit bias in the synthetic case generation task, with DeepSeek V3 following closely behind. In contrast, for the evaluation of explicit bias in differential diagnosis, the DeepSeek model outperformed all others across all metrics, emerging as the best-performing model. These findings demonstrate that DeepSeek V3 not only performs well in implicit bias detection but also demonstrates a clear advantage in scenarios involving explicit bias; by comparison, GPT-4.1 did not maintain leading performance across both tasks.

\medskip

Among all significant cases observed in Sub-experiment 2, "Asian vs Black" and "Asian vs White" were group comparisons that occurred in each of the Gemini 2.5 Pro, Llama 3.3, and Grok 3. The underlying causes for this phenomenon are consistent with those discussed in Section \ref{sec:Intro} regarding sources of racial bias in LLMs. DeepSeek V3 was identified as the model with the lowest overall measured bias across the two benchmarks used in Experiment I.

\subsection{Experiment II}
After identifying the model with the lowest degree of racial bias, this study incorporated the AI agentic workflow described in Section \ref{subsec:exp2_agentic_workflow} to reproduce the process of Sub-experiment 2 from Experiment I, with the aim of evaluating the impact of the AI agentic workflow on model bias mitigation.


\begin{table}
  \caption{Statistical Summary of Agentic Workflow}
  \label{tab:table5.3}
  \centering
  \begin{tabular}{lc}
    \toprule
    \ &  Agentic workflow\\
    \midrule
    \ Mean p-value & 0.8207\\
    \ Median p-value & 1.0000\\
    \ Bias detection rate& 0.0167\\
    \ Mean difference & 0.1639\\
  \bottomrule
\end{tabular}
\end{table}

As shown in Table \ref{tab:table5.3}, the agentic workflow based on DeepSeek V3 improved most of the reported metrics relative to the standalone DeepSeek V3 model. It produced a higher mean p-value (0.8207 versus 0.7859), a higher median p-value (1.0000 versus 0.8834), and a lower mean difference (0.1639 versus 0.2588). The bias detection rate, however, was slightly higher than that of standalone DeepSeek V3 (0.0167 versus 0.0000), although it remained lower than that of all other models except GPT-4.1. Taken together, these results suggest that the agentic workflow was associated with lower measured explicit bias on this benchmark, but the improvement was not uniform across every reported metric.

\section{Conclusion and Future Work}
\label{sec:conclusion}

This study examined racial bias in LLMs used for medical tasks through two benchmarks: synthetic patient-case generation and differential diagnosis ranking. Using the EU AI Act as a governance lens for evaluation and documentation, the study compared five widely used models across implicit and explicit bias settings. GPT-4.1 showed the smallest deviation in the synthetic case-generation task, whereas DeepSeek V3 produced the strongest overall results in the differential diagnosis task and the lowest overall measured bias across the two benchmarks. When DeepSeek V3 was embedded in an agentic workflow, most reported explicit-bias metrics improved, although the improvement was not uniform across all measures. These findings support the use of multi-metric bias evaluation in medical AI research and suggest that retrieval-based agentic workflows may reduce some forms of explicit bias under benchmark conditions.

\medskip

There remain several areas for further work. First, the explicit-bias analysis focused on rank differences in important diagnoses and did not assess whether agreement with expert diagnoses varied systematically by race. Future work could therefore add task-performance measures such as accuracy, recall, or top-\(k\) agreement. Second, the current workflow did not include a human-in-the-loop component, which may be relevant in more realistic clinical settings. Third, the workflow was applied only to the differential diagnosis task; future work should examine whether similar designs can reduce bias in synthetic case generation or other forms of medical text production.

\section{Acknowledgements}
To support transparency and traceability, the prompt templates, experimental datasets, and code pipelines used in this study are available in our GitHub repository \cite{KillerBeeH2025}.

\bibliographystyle{unsrt}  
\bibliography{references}

\end{document}